\documentclass[%
 amsmath,amssymb,
 aps,
prd,showpacs,twocolumn,superscriptaddress
]{revtex4-2} 

\usepackage{graphicx} 
\usepackage{subfigure} 
\usepackage{dcolumn} 

\usepackage{amsmath} 
\usepackage{amssymb} 
\usepackage{bm} 
\usepackage{color}

\usepackage{siunitx}
\usepackage[normalem]{ulem} 
\usepackage[dvipsnames]{xcolor} 
\usepackage{hyperref} 
\usepackage{bm} 
\usepackage{subfiles} 
\usepackage{amsthm,amsmath,amssymb} 
\usepackage{mathrsfs} 
\usepackage{url} 
\usepackage{orcidlink}

\newcommand\be{\begin{equation}}
\newcommand\ee{\end{equation}}
\newcommand\ba{\begin{eqnarray}}
\newcommand\ea{\end{eqnarray}}
\def\ba#1\ea{\begin{align}#1\end{align}}

\begin{document}

\title{A self-consistent EOB--Teukolsky framework for generic extreme mass-ratio inspirals} 
\author{Xingyu Zhong} 
\affiliation{School of Fundamental Physics and Mathematical Sciences, Hangzhou Institute for Advanced Study, UCAS, Hangzhou 310024, China}
\author{Wen-Biao Han~\orcidlink{0000-0002-2039-0726}} 
 \email{wbhan@shao.ac.cn} 
 \affiliation{Shanghai Astronomical Observatory, Shanghai, 200030, China} 
\affiliation{School of Fundamental Physics and Mathematical Sciences, Hangzhou Institute for Advanced Study, UCAS, Hangzhou 310024, China} 
\affiliation{School of Astronomy and Space Science, University of Chinese Academy of Sciences,
Beijing, 100049, China} 
\author{Ye Jiang} 
\affiliation{Shanghai Astronomical Observatory, Shanghai, 200030, China} 
\affiliation{School of Astronomy and Space Science, University of Chinese Academy of Sciences,
Beijing, 100049, China} 
\author{Chen Zhang} 
\affiliation{Shanghai University of Engineering Science, Shanghai, 201620, China} 
\author{Xaobo Zou} 
\affiliation{School of Fundamental Physics and Mathematical Sciences, Hangzhou Institute for Advanced Study, UCAS, Hangzhou 310024, China}
\date{\today} 

\begin{abstract} 


We present a full-relativistic waveform model for extreme mass-ratio inspirals (EMRIs) by self-consistently combining the effective one-body (EOB) formalism with the Teukolsky equation. The model incorporates analytical, mass-ratio-informed geodesic solutions within a deformed Kerr metric into the source term of the Teukolsky equation, establishing a direct connection between finite-mass-ratio orbital dynamics and gravitational-wave emission. The resulting frequency-domain formulation is coupled to a high-performance solver for the homogeneous Teukolsky equation, enabling rapid evaluation of the tens of thousands of modes required for accurate EMRI waveforms. We generate waveforms and radiation fluxes for generic Kerr orbits and investigate the influence of finite-mass-ratio corrections beyond the test-particle limit. The results show that mass-ratio-dependent deformations produce measurable modifications to radiation fluxes, and accumulated waveform phases over observationally relevant timescales. Our framework provides a generic-orbit EOB--Teukolsky waveform model for future space-based GW data analysis. 
\end{abstract} 

\maketitle
\section{\label{intro} Introduction} 

Currently, the ground-based gravitational-wave (GW) detector network of the LIGO--Virgo--KAGRA collaboration~\cite{LIGO, Virgo, KAGRA} has completed its fourth observing run~\cite{o4}. The milestone event GW250114~\cite{GW250114}, with an extremely high signal-to-noise ratio (SNR, $\rho$), verified Hawking's black-hole area theorem, marking the transition of GW astronomy from the era of discovery to the era of precision measurement. Meanwhile, future space-based GW missions such as LISA~\cite{lisa}, Taiji~\cite{Taiji}, and TianQin~\cite{TianQin} are progressing rapidly. In the 2030s, global GW astronomy will enter a new era of multiband, high-precision, and coordinated observations. 

Extreme mass-ratio inspirals (EMRIs)~\cite{EMRI} are one of the most important sources for space-based GW detectors. Such binary systems consist of a massive black hole and a stellar-mass compact object orbiting it with a mass ratio of $\sim 10^{-4} - 10^{-7}$. The GWs of EMRI lie within the sensitive frequency band of space-based detectors and last from months to years depending on the mass ratio. Modeling and extracting EMRI signals in space-based GW observations is of great scientific importance. As the source parameters can be determined with sub-percent precision~\cite{Babak:2017prd}, this opens up possibilities for precisely testing general relativity in the strong-field regime, inferring the evolutionary history of black hole populations at galactic centers, constraining the distribution models of dark matter, providing new independent constraints on cosmological parameters such as the Hubble constant, and exploring the equation of state of dark energy~\cite{Barausse:2007prd,Barausse:2008prd, MacLeod:2008prd, Gair:2010prd, Laghi:2021mnras}.

For EMRI events, the orbit of the secondary object lies deep within the strong-field region, rendering Post-Newtonian expansions entirely inadequate. At the same time, Numerical Relativity remains impractical due to the long duration of the inspiral phase. The only viable approach to model accurate EMRI waveforms is through Black Hole Perturbation Theory. For a typical EMRI signal, a mismatch on the order of $\sim 1/\rho^2$ introduces a systematic bias comparable to the statistical error. This implies that the orbital phase error must be controlled to within $\mathcal{O}(1)$ rad.

A standard approach to modeling EMRIs is the two-timescale self-force (SF) formalism, which systematically accounts for the small body's gravitational backreaction on a Kerr geodesic~\cite{Tanja:2008prd}. While the adiabatic and first post-adiabatic approximations are expected to satisfy the accuracy requirements of space-based detectors ~\cite{Burke:2023prd}, the computation of high-order SF terms particularly for generic orbits remains a significant challenge~\cite{Maarten:2018prd, Barry:2023prd}. 

The effective one-body (EOB) formalism provides a powerful alternative~\cite{Buonanno:1999prd}. By mapping the relativistic two-body dynamics to the motion of a test mass in a deformed Kerr metric, the EOB model calibrated against perturbation theory and numerical relativity~\cite{Pompili:2023prd, VandeMeent:2023prd} delivers a complete, semi-analytical description of the inspiral, merger, and ringdown. Its versatility makes it well-suited for systematic GW data analysis. The emitted waveform is subsequently modeled by coupling the EOB dynamics to the Teukolsky formalism, which describes curvature perturbations via the Weyl scalar $\Psi_4$; at infinity, $\Psi_4$ yields the two independent GW polarizations.

In this work, we develop a hybrid EOB--Teukolsky framework for EMRI waveform generation. This approach integrates the conservative orbital dynamics and analytic resummation of the EOB model with the rigorous perturbative foundation of the Teukolsky equation, offering a self-consistent and computationally efficient route to high-accuracy waveforms. We present the theoretical formulation, describe the numerical workflow, including a frequency-domain Teukolsky solver, and assess the model's accuracy and performance. Furthurmore, the EOB construction also provides a natural framework for exploring larger mass ratios in future work, although the present study focuses exclusively on the EMRI regime.

The paper is organized as follows. Section~\ref{sec:eob_analytical} presents the analytical EOB geodesics. Section~\ref{sec:teukolsky_waveform} introduces the frequency-domain Teukolsky framework with EOB sources. Section~\ref{sec:numerics} describes the numerical methods. Section~\ref{sec:results} presents representative waveforms and quantitative comparisons. Finally, Sec.~\ref{sec:conclusion} summarizes our findings and outlines future directions.

\section{Analytical EOB Geodesic Solutions of bound timelike orbits}
\label{sec:eob_analytical}

The foundation of our EOB-Teukolsky waveform model for EMRIs is the conservative orbital dynamics within the deformed Kerr spacetime of the effective one-body formalism. This section details the derivation of analytical solutions for bound timelike geodesics in this framework, following the work of Zhang and Han~\cite{Zhang:2025prd}. These solutions, expressed in terms of elliptic integrals using Mino time as the independent variable, are crucial for the efficient computation of orbital frequencies and the subsequent Fourier decomposition needed for the frequency-domain Teukolsky equation.

\subsection{EOB Hamiltonian}
\label{subsec:eob_hamiltonian}

We consider an EMRI  with central black hole mass $m_1$ and companion mass $m_2$, neglecting radiation reaction for the conservative dynamics. The EOB Hamiltonian maps the two-body dynamics to a test particle of reduced mass $\mu$ moving in a deformed Kerr metric, parameterized by the symmetric mass ratio $\nu = \mu / (m_1+m_2)$:
\begin{equation}
    H_{\rm EOB} = M \sqrt{1 + 2\nu (\hat{H}_{\rm eff} - 1)},
    \label{eq:heob}
\end{equation}
where $M=m_1+m_2$, $\hat{H}_{\rm eff} = H_{\rm eff}/\mu$ is the reduced effective Hamiltonian. The effective metric is defined by the deformation parameter $\nu$ and takes the form
\begin{equation}
\begin{split}
g_{\mathrm{eff}}^{\alpha\beta} P_\alpha P_\beta &= \frac{1}{r^2 + a^2 \cos^2\theta} \\
&\quad \times \left[ \Delta_r(r) P_r^2 + P_\theta^2 + \frac{1}{\sin^2\theta} \left( P_\phi + a \sin^2\theta P_t \right)^2 \right. \\
&\quad \left. - \frac{1}{\Delta_t(r)} \left( (r^2 + a^2) P_t + a P_\phi \right)^2 \right],
\end{split}
\end{equation}
where $a$ is the effective Kerr spin parameter, $P_t$, $P_r$, $P_\theta$, and $P_\phi$ are the canonical momenta conjugate to the coordinates $t$, $r$, $\theta$, and $\phi$, respectively and
\begin{equation}
\Delta_r(r) = \frac{r^2 A(u) + a^2}{D(u)}, \quad \Delta_t(r) = r^2 A(u) + a^2.
\end{equation}

The deformed Kerr metric potentials at 3PN order for the non-spinning case are given by~\cite{Buonanno:2000prd,Damour:2000prd}:
\begin{subequations}
\begin{align}
    A(u) &= 1 - 2u + 2\nu u^{3} + \left(\frac{94}{3} - \frac{41\pi^{2}}{32}\right)\nu u^{4}, \\
    D^{-1}(u) &= 1 + 6\nu u^2 + 2\nu u^3(26-3\nu),
\end{align}
\end{subequations}
where $u = M/r$. The metric itself is constructed to yield the correct Hamiltonian for a test particle, incorporating frame-dragging corrections through an adjustable parameters is given by~\cite{Barausse:2010prd}. The effective Hamiltonian $H_{\text{eff}}$ for a nonspinning particle in the deformed-Kerr metric then takes the form
\begin{equation}
\begin{split}
H_{\text{eff}} =& \frac{g^{t\phi}}{g^{tt}} P_\phi \\
&+ \frac{1}{\sqrt{-g^{tt}}} \sqrt{\mu^2 + \left[ g^{\phi\phi} - \frac{(g^{t\phi})^2}{g^{tt}} \right] P_\phi^2 + g^{rr} P_r^2 + g^{\theta\theta} P_\theta^2},
\end{split}
\end{equation}
The expression for the deformed metric is available in Ref.~\cite{Zhang:2025prd}.

\begin{widetext}
\centering
\begin{subequations}
\begin{align}
P_\theta^2 &= \mu^2\hat{Q} - \mu^2\cos^2\theta \left[ a^2 (1 - \hat{H}_\text{eff}^2) + \frac{\hat{L}_z^2}{\sin^2\theta} \right], \\
P_r^2 &= \mu^2\frac{[\, a \hat{L}_z - (r^2 + a^2) \hat{H}_\text{eff} \,]^2 - (r^2 A(u) + a^2) \Bigl[ r^2 + (a \hat{H}_\text{eff} - \hat{L}_z)^2 + \hat{Q} + 2 \frac{\tilde{\omega}_\text{fd} + a r^2 (A(u) - 1)}{r^2 A(u) + a^2} \hat{H}_\text{eff} \hat{L}_z - G(r) \hat{L}_z^2 \Bigr]}{(r^2 A(u) + a^2)^2 D^{-1}(u)}.
\end{align}
\end{subequations}
\end{widetext}

where
\begin{equation}
G(r) = \frac{\tilde{\omega}_{\text{fd}}^2 - a^2 r^4 (A(u) - 1)^2}{(r^2 A(u) + a^2)(r^2 + a^2)^2}, 
\end{equation}
and \(\hat{Q}\equiv Q/\mu\) is the semi-Carter constant given in~\cite{Zhang:2021prd}
\begin{equation}
\hat{Q} = \cos^2 \theta_{\min} \left[ a^2 \left(1 - \hat{H}_{\text{eff}}^2 \right) + \frac{\hat{L}_z^2}{\sin^2 \theta_{\min}} \right]. 
\end{equation}
where $\theta_{\min}=\pi/2-\theta_{\rm inc}$.

\subsection{Geodesic Equations in Mino Time}
\label{subsec:mino_time_geodesics}

We use Mino time $\lambda$ to separate the radial and polar librations.  The
equations of motion take the form
\begin{equation}
\label{eq:eob_mino_equations}
\begin{aligned}
 \left(\frac{dr}{d\lambda}\right)^2
 &=
 D^{-1}(u)R(r),\,\,
 \left(\frac{d\cos\theta}{d\lambda}\right)^2
 =
 \Theta(\cos\theta),
 \\
 \frac{d\phi}{d\lambda}
 &=
 V_\phi^r(r)+V_\phi^\theta(\theta),\,\,
 \frac{dt}{d\lambda}
 =
 V_t^r(r)+V_t^\theta(\theta).
\end{aligned}
\end{equation}
The radial and polar terms are
\begin{subequations}
\label{eq:V_functions}
\begin{align}
 V_t^r(r)
 &=
 \sqrt{1+2\nu(E-1)}
 \frac{E(r^2+a^2)^2-\widetilde{\omega}_{\rm fd}L_z}
      {\Delta_t},
 \\
 V_t^\theta(\theta)
 &=
 -\sqrt{1+2\nu(E-1)}\,a^2E\sin^2\theta,
 \\
 V_\phi^r(r)
 &=
 \frac{\widetilde{\omega}_{\rm fd}E-a^2L_z}
      {r^2A(u)+a^2}
 \\&-
 \frac{4a^2\nu
 \left[
 20a^2-8+
 \left(\frac{94}{3}-\frac{41\pi^2}{32}\right)u
 \right]L_z}
 {[r^2A(u)+a^2](r^2+a^2)^2},
 \\
 V_\phi^\theta(\theta)
 &=
 \frac{L_z}{\sin^2\theta}.
\end{align}
\end{subequations}
where the energy of the system is given by $E=H_{\rm EOB}$.

The exact EOB radial function $R(r)$ is replaced by a factored quartic
$\mathcal{R}(r)$ with the same physical turning points
$r_1=p/(1-e)$ and $r_2=p/(1+e)$:
\begin{equation}
\begin{split}
 \mathcal{R}(r)
 &=
 [aL_z-(r^2+a^2)E]^2
 \\&-(r^2-2Mr+a^2)\left[r^2+(aE-L_z)^2+Q\right]
 \\&-C_\mathcal{R}\nu
 \\
 &=
 (1-E^2)(r_1-r)(r-r_2)(r-r_3)(r-r_4).
\end{split}
\label{eq:R_factorized}
\end{equation}
The remaining roots are
\begin{equation}
\begin{split}
 r_3
 &=
 \frac{(A+B)+\sqrt{(A+B)^2-4AB}}{2},
\,\,
 r_4=\frac{AB}{r_3},
 \\
 A+B
 &=
 \frac{2}{1-E^2}-(r_1+r_2),
\,\,
 AB=
 \frac{a^2Q+C_\mathcal{R}\nu}{(1-E^2)r_1r_2}.
\end{split}
\end{equation}
The coefficient $C_\mathcal{R}$ is determined by matching the radial
half-period,
\begin{equation}
 \int_{r_2}^{r_1}\frac{dr}{\sqrt{\mathcal{R}(r)}}
 =
 \int_{r_2}^{r_1}\frac{dr}{\sqrt{R(r)}}.
 \label{eq:CR_matching}
\end{equation}
This is the only approximation introduced in the analytical EOB geodesic
construction with an error of order $\mathcal{O}(10^{-2}\nu)$, which is negligible for EMRIs.  It preserves the apoapsis and periapsis while allowing the radial period and the inverse map $r(w_r)$ to be evaluated analytically.

\subsection{Fundamental Frequencies}
\label{subsec:freqs_mino_time}

The periods of the radial and polar motions in Mino time, $\Lambda_r$ and $\Lambda_\theta$, are defined by integrals over one complete libration. These can be expressed analytically using elliptic integrals:
\begin{subequations}
\label{eq:Lambda_periods}
\begin{align}
 \Lambda_r
 =&
 \frac{4}{\sqrt{(1-E^2)(r_1-r_3)(r_2-r_4)}}
 \\&\left[
 K(k_r)-3\nu\frac{Z_2(\pi/2)}{r_2^2}
       -26\nu\frac{Z_3(\pi/2)}{r_2^3}
 \right],
 \\
 \Lambda_\theta
 =&
 \frac{4K(k_\theta)}{\sqrt{Q/z_-}},
\end{align}
\end{subequations}
where, $K(k)$ is the complete elliptic integral of the first kind. The moduli are $k_r^2 = \frac{r_1-r_2}{r_1-r_3}\frac{r_3-r_4}{r_2-r_4}$ and $k_\theta^2 = z_{-}/z_{+}$. The terms $Z_2$ and $Z_3$ arise from the mass-ratio corrections in the metric potential $D^{-1}(u)$ and are also given in terms of elliptic integrals.


The corresponding angular frequencies with respect to Mino time are:
\begin{equation}
    \Upsilon_r = \frac{2\pi}{\Lambda_r}, \quad \Upsilon_\theta = \frac{2\pi}{\Lambda_\theta}.
    \label{eq:Upsilon_r_theta}
\end{equation}

The average rates of change of coordinate time $t$ and azimuth $\phi$ with respect to $\lambda$ are given by the zero-frequency ($k=0, n=0$) Fourier modes $\Gamma \equiv T_{00}$ and $\Upsilon_\phi \equiv \Phi_{00}$ of $dt/d\lambda$ and $d\phi/d\lambda$ from Eqs.~(\ref{eq:eob_mino_equations}):
\begin{subequations}
\begin{align}
 \Gamma
 &=
 \frac{1}{2\pi}
 \int_0^{2\pi}V_t^r[r(w_r)]\,dw_r
 +
 \frac{1}{2\pi}
 \int_0^{2\pi}V_t^\theta[\theta(w_\theta)]\,dw_\theta,\label{eq:Gamma}
 \\
 \Upsilon_\phi
 &=
 \frac{1}{2\pi}
 \int_0^{2\pi}V_\phi^r[r(w_r)]\,dw_r
 +
 \frac{1}{2\pi}
 \int_0^{2\pi}V_\phi^\theta[\theta(w_\theta)]\,dw_\theta.\label{eq:Upsilon_phi}
\end{align}
\end{subequations}
where $w_r = \Upsilon_r \lambda$ and $w_\theta = \Upsilon_\theta \lambda$ are the angle variables for the radial and polar motions, which are $2\pi$-periodic.

The fundamental frequencies as measured by a distant observer with respect to coordinate time $t$ are obtained by combining the Mino-time frequencies with the average time dilation factor $\Gamma$:
\begin{equation}
    \Omega_r = \frac{\Upsilon_r}{\Gamma}, \quad \Omega_\theta = \frac{\Upsilon_\theta}{\Gamma}, \quad \Omega_\phi = \frac{\Upsilon_\phi}{\Gamma}.
    \label{eq:Omega_physical}
\end{equation}
These expressions provide the first semi-analytical formulas for the orbital frequencies that include leading-order mass-ratio corrections within the EOB framework.

\subsection{Analytical Orbit and Azimuthal Phase}
\label{subsec:orbital_phases}

The decoupling of motions in Mino time enables an efficient Fourier decomposition of any function $f[r(w^{r}), \theta(w^{\theta})]$ of the orbital coordinates. The Fourier coefficients with respect to $w$ are:
\begin{equation}
    \tilde{f}_{k n} = \frac{1}{(2\pi)^{2}} \int_0^{2\pi} dw^{r} \int_0^{2\pi} dw^{\theta} \, f(r,\,\theta) e^{i\left(k w^{\theta}+n w^{r}\right)}.
    \label{eq:fourier_coeff_lambda}
\end{equation}
These Fourier coefficients $\tilde{f}_{kn}$ can also be converted into the observer time $t$ Fourier coefficients $f_{kn}$ for the expansion:
\begin{equation}
    f[r(t),\theta(t)] = \sum_{k, n} f_{k n} e^{-i (k\Omega_\theta + n\Omega_r) t}.
    \label{eq:fourier_expansion_t}
\end{equation}


To evaluate the integrals in Eqs.~(\ref{eq:Gamma}), (\ref{eq:Upsilon_phi}), and (\ref{eq:fourier_coeff_lambda}), explicit expressions for $r(w_r)$ and $\cos\theta(w_\theta)$ are required. In the test-particle limit ($\nu \to 0$), these are given exactly by Jacobi elliptic functions. For the deformed Kerr case ($\nu \neq 0$), the relation $r(w_r)$ is found by solving the inverted integral from Eq.~(\ref{eq:eob_mino_equations}). The result can be written as:
\begin{subequations}
\begin{align}
    y_r &= \frac{2\mathcal{F}_3 x + \sqrt{\mathcal{F}_2^2 - 4\mathcal{F}_1\mathcal{F}_3} - \mathcal{F}_2}{2\mathcal{F}_3}, \\
    r(w_r) &= \frac{r_3 (r_1 - r_2) y_r^2 - r_2 (r_1 - r_3)}{(r_1 - r_2) y_r^2 - (r_1 - r_3)},
    \label{eq:r_wr_solution}
\end{align}
\end{subequations}
where $x = \operatorname{sn}(\varphi_r(w_r), k_r)$ is the test-particle solution, and $\mathcal{F}_{1,2,3}$ are functions of $x$, $w_r$, and $\nu$ involving elliptic integrals~\cite{Zhang:2025prd}. The polar motion remains formally identical to the Kerr case:
\begin{equation}
    \cos\theta (w_\theta) = \sqrt{z_{-}} \, \operatorname{sn}\left( \varphi_\theta(w_\theta), k_\theta \right),
    \label{eq:costheta_wtheta}
\end{equation}
where the phase function $\varphi_\theta(w_\theta)$ is a piecewise linear function of $w_\theta$. The analytical expressions for the frequencies $\Omega_{r,\theta,\phi}$ have been validated against direct numerical integration of the EOB equations of motion for a wide range of orbital parameters: eccentricity $e$, semi-latus rectum $p$, spin $a$, and inclination $\theta_{\rm inc}$. The relative difference between the analytical and numerical frequencies scales with the mass ratio $\nu$, typically lying in the range $[10^{-3}\nu, 10^{-2}\nu]$, confirming the accuracy of the approximations for EMRI applications.


With the turning-point initial conditions $t(0)=\phi(0)=0$, the oscillatory
parts of $t$ and $\phi$ are written as sine series,
\begin{subequations}
\label{eq:phase_sine_series}
\begin{align}
 t(\lambda)
 &=
 \Gamma\lambda+\Delta t_r(w_r)+\Delta t_\theta(w_\theta),
 \\
 \phi(\lambda)
 &=
 \Upsilon_\phi\lambda+\Delta\phi_r(w_r)+\Delta\phi_\theta(w_\theta),
 \end{align}
\end{subequations}

with
\begin{subequations}
\label{eq:phase_sine_series_2}
\begin{align}
 \Delta t_r(w_r)
 &=
 \sum_{n=1}^{N_r}\Delta t^r_n\sin(nw_r),
\\
 \Delta t_\theta(w_\theta)
 &=
 \sum_{k=1}^{N_\theta}\Delta t^\theta_k\sin(kw_\theta),
 \\
 \Delta\phi_r(w_r)
 &=
 \sum_{n=1}^{N_r}\Delta\phi^r_n\sin(nw_r),
\\
 \Delta\phi_\theta(w_\theta)
 &=
 \sum_{k=1}^{N_\theta}\Delta\phi^\theta_k\sin(kw_\theta).
\end{align}
\end{subequations}
The truncation orders $N_r$ and $N_\theta$ are chosen so that increasing them
does not change the source amplitudes within the target numerical tolerance.
The coefficients are computed from the cosine modes of the separated
derivatives,
\begin{subequations}
\label{eq:phase_coefficients}
\begin{align}
 \Delta t^r_n
 &=
 \frac{2}{n\pi\Upsilon_r}
 \int_0^\pi V_t^r[r(w_r)]\cos(nw_r)\,dw_r,
 \\
 \Delta t^\theta_k
 &=
 \frac{2}{k\pi\Upsilon_\theta}
 \int_0^\pi V_t^\theta[\theta(w_\theta)]\cos(kw_\theta)\,dw_\theta,
 \\
 \Delta\phi^r_n
 &=
 \frac{2}{n\pi\Upsilon_r}
 \int_0^\pi V_\phi^r[r(w_r)]\cos(nw_r)\,dw_r,
 \\
 \Delta\phi^\theta_k
 &=
 \frac{2}{k\pi\Upsilon_\theta}
 \int_0^\pi V_\phi^\theta[\theta(w_\theta)]\cos(kw_\theta)\,dw_\theta.
\end{align}
\end{subequations}
The use of the half-interval $[0,\pi]$ follows from the evenness of
$r(w_r)$, $\theta(w_\theta)$, and the separated derivatives.

\section{Teukolsky Formalism with EOB Source}
\label{sec:teukolsky_waveform}

The gravitational waveform is computed by solving the frequency-domain
Teukolsky equation on the Kerr background of the primary black hole.  The EOB
deformation is inserted through the particle worldline and its four-velocity
in the source term; the Teukolsky differential operator and its separability
are not modified.  In the presence of a point-particle source,
\begin{equation}
 \mathcal{O}[\psi_4]=\mathcal{T}[z(\lambda),\dot{z}(\lambda)] ,
 \label{eq:teukolsky_inhomog}
\end{equation}
where $\mathcal{O}$ is the spin $-2$ Teukolsky operator and
$\mathcal{T}$ is constructed from the stress-energy tensor of the secondary.

\subsection{Teukolsky formalism}
We use the separated expansion
\begin{equation}
 \psi_4
 =
 \rho^4
 \int d\omega
 \sum_{\ell m}
 R_{\ell m\omega}(r)\,
 {}_{-2}S_{\ell m\omega}(\theta)\,
 e^{-i\omega t+im\phi},
 \label{eq:psi4_expansion}
\end{equation}
where $\rho=-(r-ia\cos\theta)^{-1}$.  The angular functions satisfy
\begin{align}
&\left[
\frac{1}{\sin\theta}\frac{d}{d\theta}
\left(\sin\theta\frac{d}{d\theta}\right)
-a^2\omega^2\sin^2\theta
-\frac{(m-2\cos\theta)^2}{\sin^2\theta}
\right.
\notag\\
&\left.
\qquad
+4a\omega\cos\theta-2+\lambda_{\ell m}(\omega)
\right]
{}_{-2}S_{\ell m\omega}(\theta)=0 .
\label{eq:angular_eq}
\end{align}

The radial Teukolsky equation is
solved as an ordinary differential equation,
\begin{equation}
 \Delta^2\frac{d}{dr}
 \left(
 \frac{1}{\Delta}\frac{dR_{\ell m\omega}}{dr}
 \right)
 -\mathcal{V}_{\ell m}(r,\omega)R_{\ell m\omega}
 =
 -\mathfrak{T}_{\ell m\omega}(r),
 \label{eq:radial_inhomogeneous}
\end{equation}
where $\Delta = r^2 - 2Mr + a^2$, $\mathcal{V}_{lm}$ is a complex potential, and $\mathfrak{T}_{l m \omega}(r)$ is the source term, projected onto the spheroidal harmonic and integrated over $\theta$ and $\lambda$. 

Let $R^{\infty}_{\ell m\omega}$ and $R^H_{\ell m\omega}$ denote the two homogeneous solutions satisfying outgoing boundary conditions at infinity and ingoing boundary conditions at the horizon, respectively.  The physical solution has
the asymptotic form
\begin{subequations}
\begin{align}
 R_{\ell m\omega}(r\to\infty)
 &=
 Z^\infty_{\ell m\omega}\,
 r^3e^{i\omega_{m\omega}r^*}
 +\mathcal{O}(1/r),
 \label{eq:radial_inf_amp}
 \\
 R_{\ell m\omega}(r\to r_+)
 &=
 Z^H_{\ell m\omega}\,
 \Delta^2e^{-i\tilde{\omega}_{m\omega}r^*}
 +\mathcal{O}(\Delta^2),
 \label{eq:radial_hor_amp}
\end{align}
\end{subequations}
where $r^*$ is the tortoise coordinate,
$r_+=M+\sqrt{M^2-a^2}$, and
$\tilde{\omega}_{m\omega}=\omega_{m\omega}-m\Omega_H$ with
$\Omega_H=a/(2Mr_+)$.  Equivalently, the variation-of-parameters expression
for the amplitudes is
\begin{subequations}
\begin{align}
 Z^\infty_{\ell m\omega}
 &=
 \frac{1}{W_{\ell m\omega}}
 \int_{r_+}^{\infty}dr'\,
 \frac{R^H_{\ell m\omega}(r')}{\Delta'^2}
 \mathfrak{T}_{\ell m\omega}(r'),
 \label{eq:Zinf_def}
 \\
 Z^H_{\ell m\omega}
 &=
 \frac{1}{W_{\ell m\omega}}
 \int_{r_+}^{\infty}dr'\,
 \frac{R^\infty_{\ell m\omega}(r')}{\Delta'^2}
 \mathfrak{T}_{\ell m\omega}(r'),
 \label{eq:ZH_def}
\end{align}
\end{subequations}
where $W_{\ell m\omega}$ is the Wronskian.  



The bridge between the EOB orbit and the Teukolsky amplitudes is obtained by rewriting the source integral in Mino time.  After the angular projection and the radial delta-function manipulations, the source amplitude can be written schematically as:
\begin{equation}
 Z^{H,\infty}_{\ell m}(r,\omega)
 =
 \int dt\,
 e^{i\omega t-im\phi(t)}
 I^{H,\infty}_{\ell m}(t,r,\omega),
 \label{eq:Zlm_source_t}
\end{equation}
the factor
$I^{H,\infty}_{\ell m}$ contains the tetrad projections of the source, the spin-weighted spheroidal harmonics and their derivatives, and the homogeneous radial function.

Substituting
Eqs.~\eqref{eq:phase_sine_series} into Eq.~\eqref{eq:Zlm_source_t} and
changing variables from $t$ to $\lambda$ gives
\begin{equation}
 Z^{H,\infty}_{\ell m}(r,\omega)
 =
 \int d\lambda\,
 e^{i(\omega\Gamma-m\Upsilon_\phi)\lambda}
 J^{H,\infty}_{\ell m}(\lambda,r,\omega),
 \label{eq:Zlm_source_lambda}
\end{equation}
with
\begin{align}
 J^{H,\infty}_{\ell m}(\lambda,r,\omega)
 &=
 \frac{dt}{d\lambda}
 I^{H,\infty}_{\ell m}(\lambda,r,\omega)
 \exp\{i\omega[\Delta t_r(w_r)+\Delta t_\theta(w_\theta)]
 \notag\\
 &\qquad
 -im[\Delta\phi_r(w_r)+\Delta\phi_\theta(w_\theta)]\}.
 \label{eq:Jlm_biperiodic}
\end{align}
All nontrivial $\lambda$ dependence in $J^{H,\infty}_{\ell m}$ is through the
periodic variables $(w_r,w_\theta)$, so it is biperiodic.  We expand it as
\begin{equation}
 J^{H,\infty}_{\ell m}(\lambda,r,\omega)
 =
 \sum_{kn}
 J^{H,\infty}_{\ell mkn}(r,\omega)
 e^{-i(k\Upsilon_\theta+n\Upsilon_r)\lambda}.
 \label{eq:Jlm_fourier}
\end{equation}
The $\lambda$ integral then collapses to a discrete spectrum,
\begin{equation}
 \omega_{mkn}
 =
 m\Omega_\phi+k\Omega_\theta+n\Omega_r .
 \label{eq:mode_frequencies}
\end{equation}
The corresponding source amplitude is
\begin{equation}
\begin{aligned}
 Z^{H,\infty}_{\ell mkn}
 =&
 \mathcal{B}^{H,\infty}_{\ell mkn}
 \frac{1}{2\pi\Gamma}
 \int_0^{2\pi}dw_\theta
 \int_0^{2\pi}dw_r\,
 I^{H,\infty}_{\ell m}
 \frac{dt}{d\lambda}
 \\
 &\times
 \exp\left\{
 i\left[
 \omega_{mkn}(\Delta t_r+\Delta t_\theta)
 \right.\right. 
 \\
 &\left.\left. 
 -m(\Delta\phi_r+\Delta\phi_\theta)
 +kw_\theta+nw_r
 \right]
 \right\} 
 \label{eq:eob_teukolsky_amplitudes}
\end{aligned}
\end{equation}
The prefactors $\mathcal{B}^{H,\infty}_{\ell mkn}$ convert the local homogeneous solutions to the physical horizon and infinity amplitudes. 
Thus the EOB deformation enters the Teukolsky source through $r(w_r)$, $\theta(w_\theta)$, $P_r(w_r)$, $P_\theta(w_\theta)$, the phase corrections $\Delta t$ and $\Delta\phi$, and the frequencies
$(\Omega_r,\Omega_\theta,\Omega_\phi)$, while the Teukolsky mode structure remains the standard $(\ell,m,k,n)$ decomposition.

\subsection{Computing Homogeneous Solutions by JH Method}
\label{subsec:jh_method}
The core computational task in evaluating Eqs.~(\ref{eq:Zinf_def}) and (\ref{eq:ZH_def}) is the efficient and accurate calculation of the homogeneous radial solutions $R^{\infty}_{l m k n}(r)$ and $R^{H}_{l m k n}(r)$, and their $\mathcal{B}_{\ell mkn}$, for a very large number of modes $(l, m, k, n)$. The performance of this step is the primary bottleneck in frequency-domain Teukolsky-based waveform generation for EMRIs.

The Mano-Suzuki-Takasugi (MST) method provides global analytic solutions in terms of hypergeometric functions but becomes inefficient at high frequencies and often requires arbitrary-precision arithmetic for stability~\cite{Mano:1996PTP}. The generalized Sasaki-Nakamura (SN) method transforms the equation to a form with a shorter-range potential suitable for numerical integration, but the integration process itself can be slow and suffers from numerical stiffness, especially for large $\omega$ or near-extremal spins~\cite{Hughes:2000prd}.

To overcome these limitations, our EOB-Teukolsky model employs the recently developed JH method~\cite{Jiang:2026prd} for computing the homogeneous Teukolsky solutions. The JH method is a high-performance, high-accuracy algorithm based on matching local analytic series expansions, designed specifically for the high-throughput requirements of EMRI waveform generation.

The JH method reformulates the radial Teukolsky equation and constructs its global solution by piecing together different types of local expansions:
\begin{enumerate}
    \item Local solution at the horizon: Uses Frobenius series expansions to represent the ingoing  and outgoing solutions near the event horizon .
    \item Local solution at infinity: Employs asymptotic series expansions to represent the ingoing and outgoing solutions at large radii.
    \item Bridging with ordinary point expansion: In the intermediate region, ordinary point power series expansions are used to analytically continue the solutions from the horizon or infinity, effectively ``bridging'' the domain.
    \item Global matching and amplitude extraction: The local solutions are matched at a chosen point to determine the connection coefficients between the horizon and infinity bases, as detailed in Eqs.~(35)-(45) and (50)-(54) of Ref.~\cite{Jiang:2026prd}. This directly yields the asymptotic amplitudes $B^{\text{trans}}$, $B^{\text{inc}}$, $C^{\text{trans}}$, etc., which are proportional to the $Z^{\infty, H}_{l m k n}$ amplitudes in our formalism.
\end{enumerate}

A key strength of the JH method is its handling of extreme parameters. For very small frequencies ($|\omega| \lesssim 0.01$), it switches to an expansion in confluent hypergeometric functions. For large frequencies, it employs a large frequency expansion, which reformulates the equation into a standard WKB-like form, thereby avoiding the numerical oscillation problems that plague integration-based methods.

The JH method replaces the numerical integration process of the SN method with analytic series operations, leading to a dramatic increase in speed. The JH method outperforms both the MST and SN methods by orders of magnitude:

Initialization Time ($t_i$): The JH method requires only $\mathcal{O}(10-50) \mu s$ to pre-compute coefficients for a given $(l, m, \omega)$ mode, compared to $\mathcal{O}(10-100) ms$ for the other methods.

Per-point Evaluation Time ($\bar{t}_c$): Evaluating the homogeneous solution and its derivative at a single radial point $r$ takes the JH method approximately $100-150 ns$, which is about 100 times faster than the SN method and over 10,000 times faster than the MST method for moderate frequencies.

Furthermore, the method maintains very high accuracy, with relative residuals of the solution typically below $10^{-13}$ and Wronskian deviations below $10^{-12}$ across a wide range of frequencies and black hole spins.

\subsection{Waveform and Fluxes}
\label{subsec:waveform_fluxes_implementation}

The asymptotic amplitudes $Z^{\infty}_{l m k n}$ directly yield the waveform at infinity. The relation between $\psi_4$ and the two GW polarizations $h_+$ and $h_\times$ for a distant observer is
\begin{equation}
    \psi_4(r\to\infty, t, \theta, \phi) = \frac{1}{2} \frac{\partial^2}{\partial t^2} \left( h_+ - i h_\times \right).
    \label{eq:psi4_to_h}
\end{equation}
Substituting the asymptotic form of Eq.~(\ref{eq:psi4_expansion}) with Eq.~(\ref{eq:radial_inf_amp}) and integrating twice in time (discarding static terms) gives the complete frequency-domain waveform:
\begin{equation}
    h_{+} - i h_{\times} = -\frac{2}{r} \sum_{l, m, k, n} \frac{Z^{\infty}_{l m k n}}{\omega_{m k n}^2} \, _{-2}S_{l m k n}(\theta) \, e^{-i \omega_{m k n} (t - r^*) + i m \phi}.
    \label{eq:full_waveform}
\end{equation}
The sum is over all spheroidal harmonic modes $(l, m)$ and orbital harmonics $(k, n)$. In practice, the sum must be truncated at some $l_{\text{max}}$ and $(k, n)_{\text{max}}$; convergence studies ensure the truncated result is sufficiently accurate.

\begin{figure*}[!ht]
\centering
\begin{tabular}{c}
\includegraphics[width=\linewidth]{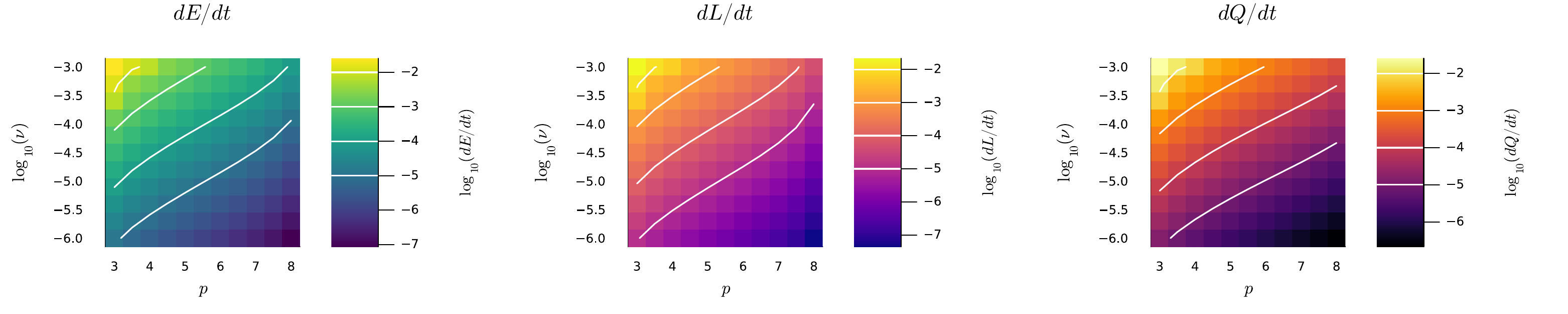}
\end{tabular}
\caption{Relative differences in the orbit-averaged fluxes $\langle dE/dt\rangle$, $\langle dL_z/dt\rangle$, and $\langle dQ/dt\rangle$ between the EOB--Teukolsky calculation and the test-particle limit. The horizontal and vertical axes show the semi-latus rectum $p$ and symmetric mass ratio $\nu$, respectively. The color scale gives $\log_{10}|\Delta\mathcal{F}|$, with $\Delta\mathcal{F}$ defined in Eq.~\eqref{eq:relative_flux_difference}. White curves mark contours of constant relative difference, $10^{-2}$, $10^{-3}$, $10^{-4}$, and $10^{-5}$.}
\label{fig:dflux3}
\end{figure*}

The total dissipative evolution for the constants of motion $(E,\, L_z,\, Q)$ is driven by the sum of the infinity and horizon fluxes, which can be expressed in terms of the Teukolsky amplitudes $Z^{\infty, H}_{l m k n}$:
\begin{subequations}
\begin{align}
    \left( \frac{dE}{dt} \right)^{\infty} &= \sum_{l, m, k, n} \frac{1}{4\pi \omega_{m k n}^2} \left| Z^{\infty}_{l m k n} \right|^2, \label{eq:E_flux_inf} \\
    \left( \frac{dE}{dt} \right)^{H} &= \sum_{l, m, k, n} \frac{\alpha_{l m k n}}{4\pi \omega_{m k n}^2} \left| Z^{H}_{l m k n} \right|^2, \label{eq:E_flux_hor} \\
    \left( \frac{dL_z}{dt} \right)^{\infty} &= \sum_{l, m, k, n} \frac{m}{4\pi \omega_{m k n}^3} \left| Z^{\infty}_{l m k n} \right|^2,  \\
    \left( \frac{dL_z}{dt} \right)^{H} &= \sum_{l, m, k, n} \frac{m \alpha_{l m k n}}{4\pi \omega_{m k n}^3} \left| Z^{H}_{l m k n} \right|^2,\\
    \left( \frac{dQ}{dt} \right)^{\infty} &= \sum_{lmkn} |Z_{lmkn}^{\infty}|^2 \times \frac{(\mathcal{L}_{mkn} + k\Upsilon_{\theta})}{2\pi\omega_{mkn}^3}, \\
\left( \frac{dQ}{dt} \right)^{\mathrm{H}} &= \sum_{lmkn} \alpha_{lmkn} |Z_{lmkn}^{\mathrm{H}}|^2 \times \frac{(\mathcal{L}_{mkn} + k\Upsilon_{\theta})}{2\pi\omega_{mkn}^3}.\label{eq:Q_flux_H} 
\end{align}
\end{subequations}
where, $\mathcal{L}_{mkn}$ and $\alpha_{l m k n}$ is
\begin{subequations}
\begin{align}
\mathcal{L}_{mkn} =& m\langle\cot^2\theta\rangle L_z - a^2\omega_{mkn}\langle\cos^2\theta\rangle E, \\
\alpha_{lmkn} =& \{256(2Mr_+)^5(\omega_{mkn} - m\Omega_{\mathrm{H}})[(\omega_{mkn} - m\Omega_{\mathrm{H}})^2 \notag\\&+ 4\epsilon^2][(\omega_{mkn} - m\Omega_{\mathrm{H}})^2 + 16\epsilon^2]\omega_{mkn}^3\}/{|C_{lmkn}|^2}.
\end{align}
\label{eq:L_alpha}
\end{subequations}

where terms $\langle\cot^2\theta\rangle$ and $\langle\cos^2\theta\rangle$ mean $\cot^2\theta$ and $\cos^2\theta$ evaluated at the $\theta$ coordinate along and then averaged, $\epsilon = \sqrt{M^2 - a^2}/{4Mr_+}$ and the term $|C_{lmkn}|^2$ is given by

\begin{equation}
\begin{aligned}
|C_{lmkn}|^2 =& [(\lambda_{lmkn}^2 + 2)^2 + 4am\omega_{mkn} - 4a^2\omega_{mkn}^2](\lambda_{lmkn}^2 \\&+ 36am\omega_{mkn} - 36a^2\omega_{mkn}^2) \\
& + (2\lambda_{lmkn} + 3)(96a^2\omega_{mkn}^2 - 48am\omega_{mkn}) \\&+ 144\omega_{mkn}^2(M^2 - a^2).
\end{aligned}
\label{eq:C_epsilon}
\end{equation}
Fig.~\ref{fig:dflux3} gives a global view of the finite-mass-ratio corrections to the orbit-averaged fluxes. For each flux $\mathcal{F}\in\{\langle dE/dt\rangle,\langle dL_z/dt\rangle,\langle dQ/dt\rangle\}$, we define
\begin{equation}
 \Delta\mathcal{F}[\%]
 =
 100\times
 \frac{\mathcal{F}_{\rm EOB\text{-}Teuk}-\mathcal{F}_{\rm test}}
 {\mathcal{F}_{\rm test}}.
 \label{eq:relative_flux_difference}
\end{equation}
The correction increases as the orbit approaches the strong-field region and as the mass ratio increases.

In our EOB-Teukolsky model, the workflow for generating adiabatic inspiral waveforms is listed as follows:
\begin{enumerate}
    \item For a given set of orbital parameters $(p, e, \theta_{\text{inc}})$, mass-ratio $\nu$ and the Kerr parameter $a$ at an instant, compute the fundamental frequencies $\Omega_{r,\theta,\phi}$ and the Fourier representations of $r(w_r)$ and $\theta(w_\theta)$ using the analytical framework of Sec.~\ref{sec:eob_analytical}.
    \item Use the EOB orbital functions to construct the mode decomposition of the source term $\mathfrak{T}_{l m k n}(r)$ via integrals over $w_r$ and $w_\theta$~\cite{Hughes:2006prd}.
    \item For each mode $(l, m, k, n)$,
    \begin{enumerate}
        \item Compute the mode frequency $\omega_{m k n}$ via Eq.~(\ref{eq:mode_frequencies}).
        \item Use the JH method to efficiently compute the homogeneous solutions $R^{\infty}_{l m k n}(r)$ and $R^{H}_{l m k n}(r)$ and their Wronskian $W_{l m k n}$. 
        \item Perform the integrals in Eqs.~(\ref{eq:Zinf_def}) and (\ref{eq:ZH_def}) to obtain the amplitudes $Z^{\infty, H}_{l m k n}$.
    \end{enumerate}
    \item Construct the waveform using Eq.~(\ref{eq:full_waveform}) and compute the flux sums in Eqs.~(\ref{eq:E_flux_inf}--\ref{eq:Q_flux_H} ). The infinite sums are truncated adaptively: the $l$-sum terminates when the current $l$-mode contribution drops below $\varepsilon_{\text{flux}}$ times the maximum contribution across all $l$-modes; the $k$-sum terminates after $B=2$ consecutive harmonic contributions fall below $\varepsilon_{\text{flux}}$ times the peak amplitude; and the $n$-sum terminates after $B=5$ consecutive contributions meet the same threshold. This ensures the dominant ``voices'' are captured. The fluxes determine the instantaneous rates of change $(dE/dt, dL_z/dt, dQ/dt)$.
    \item Use the fluxes to evolve the orbital parameters $(p, e, \theta_{\text{inc}})$ over a time step $\Delta t$ using flux-balance laws. Return to step 1 with the new parameters, iterating to produce the full inspiral and its accompanying waveform.
\end{enumerate}
The integration of the analytic EOB orbital input with the highly efficient JH solver for the perturbation equations creates a powerful synergy. The modular approach cleanly separates the conservative EOB dynamics from the dissipative Teukolsky radiation reaction. Critically, the computational efficiency of the JH method, which provides homogeneous solutions several orders of magnitude faster than previous methods, makes the generation of long, high-accuracy EMRI waveform templates through this self-consistent EOB-Teukolsky approach computationally feasible for the first time. This addresses the primary performance bottleneck and fulfills the requirements of the two-timescale approximation for practical data analysis applications with future space-based observatories.

\section{Numerical Workflow of the EOB-Teukolsky Model}
\label{sec:numerics}

The computational pipeline of the EOB-Teukolsky model integrates the analytical orbital machinery from Sec.~\ref{sec:eob_analytical} with the frequency-domain Teukolsky formalism of Sec.~\ref{sec:teukolsky_waveform}. This section details the numerical workflow and convergence strategy.

\subsection{Computational Workflow}
\label{subsec:workflow}
The generation of a waveform or flux for a given orbital snapshot $(p, e, \theta_{\text{inc}}, \nu, a)$ follows a modular sequence:
\begin{enumerate}
    \item  Compute the orbital constants $(E, L_z, Q)$ and the fundamental Mino-time frequencies $(\Upsilon_r, \Upsilon_\theta, \Gamma, \Upsilon_\phi)$ using the analytic formulas in Sec.~\ref{subsec:freqs_mino_time}. Construct the Fourier representations of $r(w_r)$, $\theta(w_\theta)$, $\Delta t(w_r, w_\theta)$, and $\Delta \phi(w_r, w_\theta)$.
    \item  For a specific $(l, m, k, n)$ mode, evaluate the source integrals for $Z^{\infty}_{l m k n}$ and $Z^{H}_{l m k n}$ as defined in Eqs.~(\ref{eq:Zinf_def}) and (\ref{eq:ZH_def}). The integrand requires the EOB orbital functions and their derivatives. The double integral over the angle variables $w_r$ and $w_\theta$ is performed using a tensor-product Clenshaw-Curtis quadrature rule, with a variable transformation $\psi \to \zeta$ to handle the large dynamic range in the radial integral for highly eccentric orbits.
    \item  Compute the homogeneous radial solutions $R^{\infty}_{l m k n}(r)$ and $R^{H}_{l m k n}(r)$ and their Wronskian $W_{l m k n}$ for the frequency $\omega_{m k n}$. We employ the highly accurate and efficient JH method to solve the homogeneous equation, which is then transformed to the Teukolsky solutions. 
    \item  Sum the contributions of all $(l, m, k, n)$ modes to construct the waveform (\ref{eq:full_waveform}) or fluxes. The infinite sums are truncated adaptively. The sum over $l$ is terminated when the contribution of an $l$-mode, $F_l$, satisfies $|F_l| < \varepsilon \cdot \max_{l'}|F_{l'}|$, where $\varepsilon$ is the target fractional accuracy (we use $\varepsilon=10^{-6}$). The sums over $k$ and $n$ are truncated using a buffer rule: we stop when $B$ consecutive harmonic contributions fall below a threshold relative to the peak mode amplitude encountered, ensuring the dominant voices of the orbit are captured.
\end{enumerate}

\subsection{Validation}
\label{subsec:validation}
The framework is validated against established test-particle results in the limit $\nu=0$. For circular equatorial orbits ($e=0$, $\theta_{\rm inc}=0^\circ$), the fluxes and waveforms agree with the frequency-domain reference calculation to within the reported accuracy ($\sim10^{-10}$). For eccentric equatorial orbits in the limit $\theta_{\rm inc}\to0$, our energy and angular-momentum fluxes agree with the results of Glampedakis and Kennefick~\cite{Glampedakis:2002prd}, typically to within $10^{-8}$ or better for the dominant fluxes at infinity. For generic Kerr orbits, our fluxes agree with the results of Fujita~\cite{Fujita:2009ptp} to within $10^{-6}$--$10^{-7}$. These tests validate the coupling of generic EOB geodesics to the Teukolsky equation in the limit of $\nu=0$.

\begin{figure*}[!ht]
\centering
\begin{tabular}{c}
\includegraphics[width=\linewidth]{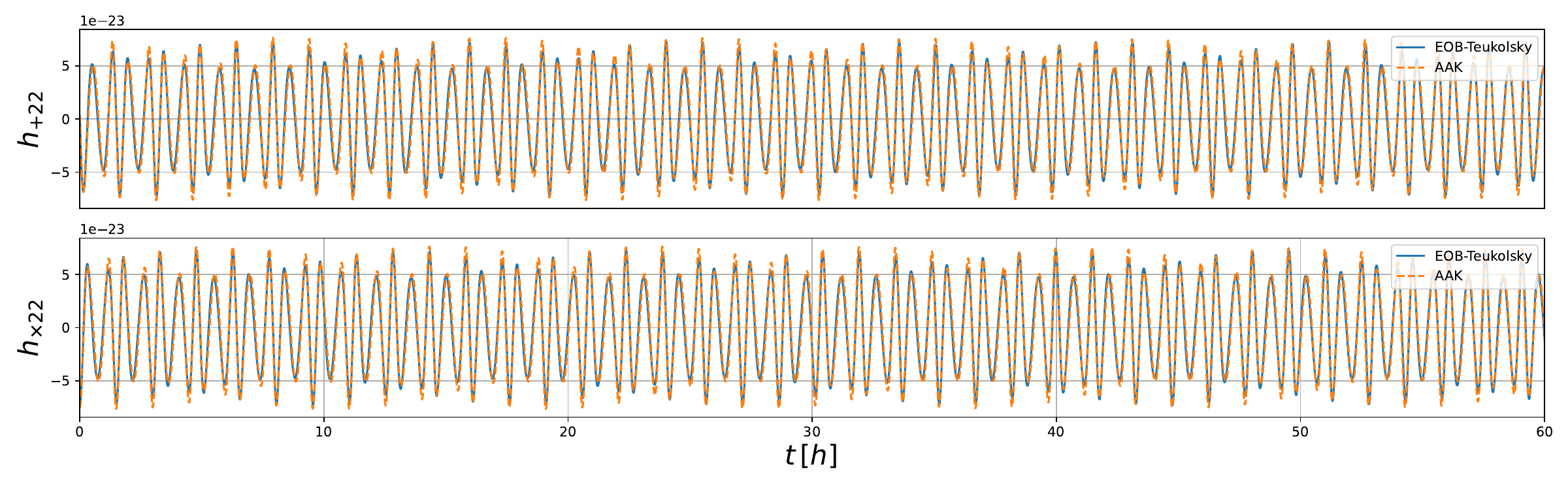}
\end{tabular}
    \caption{Comparison between the EOB--Teukolsky and AAK waveforms for $m_1=1\times10^6M_{\odot}$, $m_1=10M_{\odot}$, $p=7\,M$, $e=0.3$, $\theta_{\rm inc}=0.3$, and $a=0.9M$. }
    \label{fig:waveform_aak}
\end{figure*}

Though it is not nessasary, we compares our EOB--Teukolsky waveform with an augmented analytic kludge (AAK) waveform for the same initial orbital parameters in Fig.~\ref{fig:waveform_aak}. Be careful that the AAK just uses the Kerr geodesic and Post-Newtonian fluxes and waveforms, we can still see that the two waveforms concide each other very well in the first two days. However, with the inspiral time growing, the dephasing between these two waveforms becomes larger and larger, this will be demonstated by mismatch in the next section. 

\section{Results}
\label{sec:results}

We now assess the physical impact of the mass-ratio-dependent EOB deformation on the emitted radiation. We first examine the morphology of a representative waveform and its accumulated dephasing relative to the test-particle limit. We then quantify the corresponding changes in the orbit-averaged fluxes across the parameter space and study their dependence on the symmetric mass ratio, eccentricity, and inclination. Finally, we present noise-weighted matches for space-based detectors and summarize the convergence of the mode sums.



\begin{figure}[!ht]
\centering
\begin{tabular}{c}
\includegraphics[width=\linewidth]{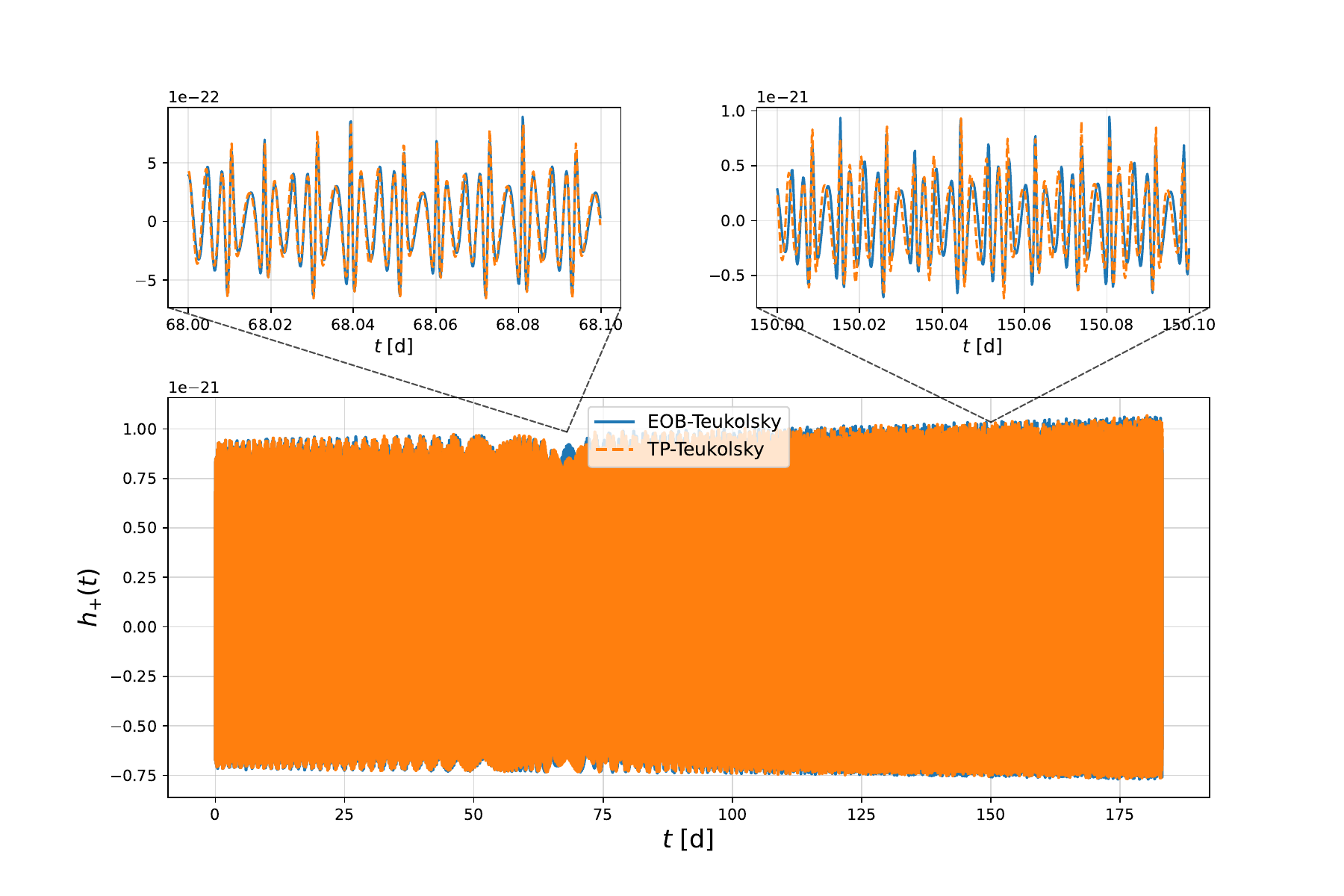}
\end{tabular}
\caption{Comparison between the EOB--Teukolsky waveform and its test-particle counterpart for the same initial orbit, $m_1=1\times10^6M_{\odot}$, $m_1=10M_{\odot}$, $p=7\,M$, $e=0.3$, $\theta_{\rm inc}=0.3$, and $a=0.9M$. The insets enlarge two late-time intervals and show the accumulated phase difference generated by the mass-ratio-dependent EOB deformation.}
\label{fig:waveform_phi}
\end{figure}

To isolate the effect of the EOB deformation, Fig.~\ref{fig:waveform_phi} compares the EOB--Teukolsky waveform with the corresponding test-particle Teukolsky waveform, obtained by setting $\nu=0$ while keeping the initial orbital parameters fixed. The two signals are initially aligned. Their relative phase subsequently drifts because the EOB deformation shifts the fundamental frequencies and therefore changes the accumulated orbital phases. The enlarged intervals near $t=68~{\rm d}$ and $t=150~{\rm d}$ make this progressive dephasing visible. The disagreement at late times is therefore the expected secular consequence of introducing mass-ratio information into the conservative orbital dynamics. However, the purpose of this comparison is not to assess waveform accuracy against an exact finite-mass-ratio solution, which is presently unavailable for generic EMRIs, but to quantify the cumulative impact of finite-mass-ratio conservative corrections introduced by the EOB dynamics.

\begin{figure}[h]
    \centering
    \includegraphics[width=\linewidth]{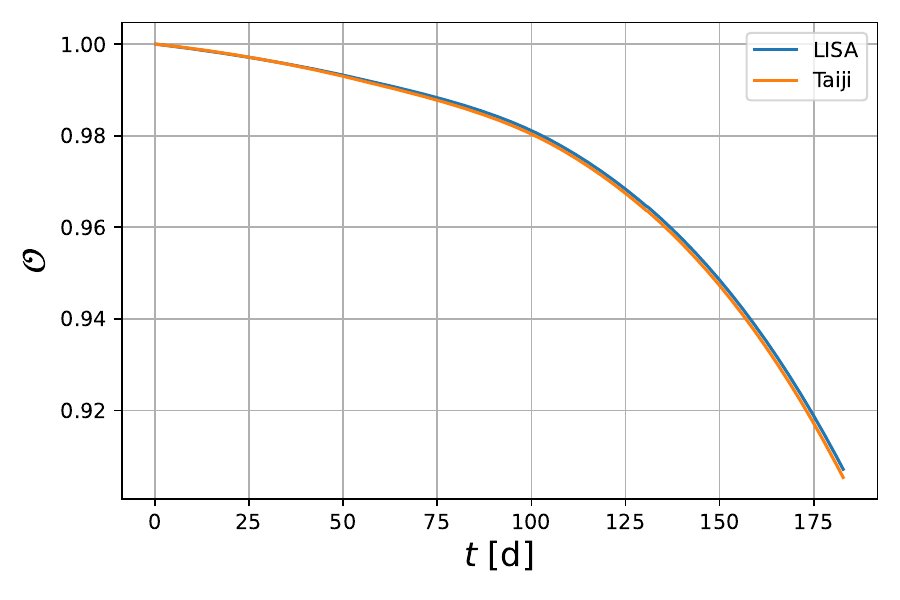}
    \includegraphics[width=\linewidth]{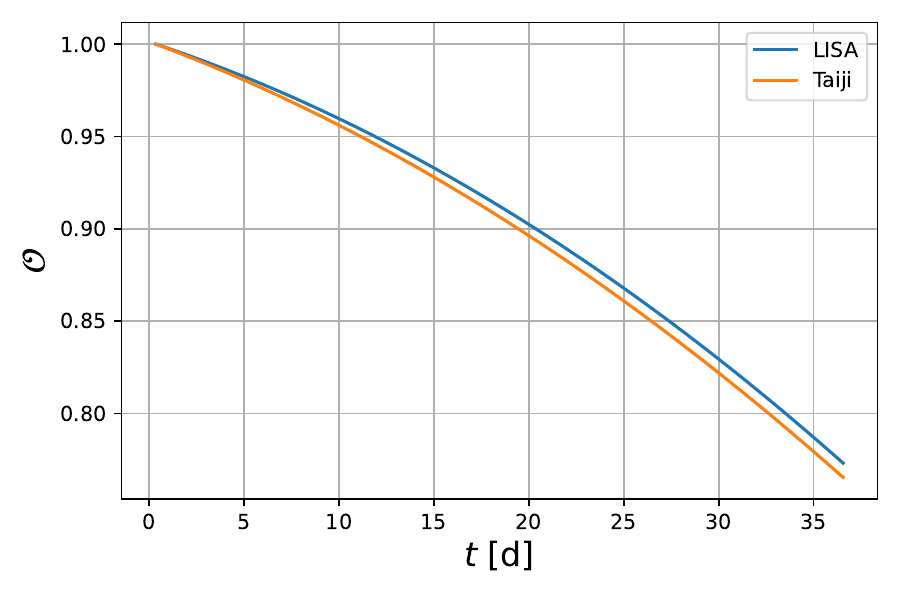}
    \caption{Accumulated detector-noise-weighted match between the EOB--Teukolsky with test-particle (top panel) and AAK waveforms (bottom panel) for $m_1=1\times10^6M_{\odot}$, $m_1=10M_{\odot}$, $p=7\,M$, $e=0.3$, $\theta_{\rm inc}=0.3$, and $a=0.9M$. Results are shown using the LISA and Taiji noise spectral densities.}
    \label{fig:waveform_o}
\end{figure}


The flux differences are small instantaneously, but their secular accumulation can become important over an observationally relevant duration. To quantify this effect, we compute the detector-noise-weighted match
\begin{equation}
 \mathcal{O}
 =
 \max_{\Delta t,\Delta\phi}
 \frac{
 (h_{\rm EOB\text{-}Teuk}|h_{\rm test})
 }{
 \sqrt{
 (h_{\rm EOB\text{-}Teuk}|h_{\rm EOB\text{-}Teuk})
 (h_{\rm test}|h_{\rm test})
 }
 } ,
 \label{eq:waveform_match}
\end{equation}
where the maximization removes an overall time shift and phase shift, and $(\,\cdot\,|\,\cdot\,)$ denotes the standard inner product weighted by the one-sided detector noise spectral density.

Fig.~\ref{fig:waveform_o} shows the accumulated match between the EOB--Teukolsky with the test-particle and AAK waveforms for the same initial orbit as in Fig.~\ref{fig:waveform_phi}. For the test partical case, the match decreases smoothly as the frequency shifts accumulate, reaching approximately $0.98$ after $100~{\rm d}$ and approximately $0.91$ near the end of the displayed interval. However, due the the AAK not just using the Kerr geodeisc but also the PN fluxes and waveforms, the mismatch increase much faster, after one month evolution, the match is around 0.8. The LISA and Taiji curves remain close because the loss of agreement is driven primarily by the intrinsic phase evolution; their slightly different noise spectra only weakly modify the weighting of the signal harmonics.

Tables~\ref{tab:flux_comparison}--\ref{tab:flux_comparison3} complement Fig.~\ref{fig:dflux3} by fixing $p=6M$ and $a=0.9M$ and varying $e$ and $\theta_{\rm inc}$. Increasing $\nu$ by one order of magnitude increases the relative flux differences by approximately one order of magnitude throughout the sampled domain. This near-linear scaling is consistent with the fact that the current EOB deformation enters at leading order in $\nu$.

At fixed mass ratio, eccentricity produces the clearest variation. For example, at $\nu=10^{-3}$ and $\theta_{\rm inc}=60^\circ$, the relative energy-flux difference grows from $0.0425\%$ at $e=0.1$ to $0.5345\%$ at $e=0.7$. The inclination dependence is weaker but becomes more visible at larger eccentricity. At $\nu=10^{-3}$ and $e=0.7$, increasing $\theta_{\rm inc}$ from $20^\circ$ to $60^\circ$ changes the relative energy-flux difference from $0.4322\%$ to $0.5345\%$. These trends indicate that mass-ratio corrections are most relevant for eccentric strong-field orbits, while inclination provides a smaller but non-negligible modulation.

The harmonic index $n$ of the peak amplitude in the flux spectrum scales approximately as $n_{\text{peak}} \sim \exp[1/2 - (3/2)\ln(1-e)]$, a scaling inherited from the Peters-Mathews formula for Keplerian orbits. The numerical results confirm that this scaling remains a useful rule of thumb even for strong-field Kerr orbits, guiding the efficient truncation of the radial harmonic sum.
\begin{table*}[t]
\caption{\label{tab:flux_comparison} Percentage differences in the orbit-averaged fluxes between the EOB--Teukolsky model with $\nu=10^{-5}$ and the test-particle limit ($\nu=0$) for bound orbits with $p=6M$, $a=0.9M$, and varying eccentricity and inclination. The differences are defined by Eq.~\eqref{eq:relative_flux_difference}.}
\centering
\begin{ruledtabular}
\begin{tabular}{ccS[table-format=1.3]S[table-format=1.3]S[table-format=1.3]}
$e$ & $\theta_{\text{inc}}$ & {$\Delta\langle dE/dt \rangle\%$} & {$\Delta\langle dL_z/dt \rangle\%$} & {$\Delta\langle dQ/dt \rangle\%$} \\ 
\hline
0.1 & $20^\circ$ & 0.0004 & 0.0003 & 0.0006 \\
0.1 & $60^\circ$ & 0.0004 & 0.0003 & 0.0006 \\
0.3 & $20^\circ$ & 0.0010 & 0.0008 & 0.0011 \\
0.3 & $60^\circ$ & 0.0012 & 0.0009 & 0.0012 \\
0.5 & $20^\circ$ & 0.0022 & 0.0017 & 0.0021 \\
0.5 & $60^\circ$ & 0.0025 & 0.0020 & 0.0024 \\
0.7 & $20^\circ$ & 0.0037 & 0.0031 & 0.0037 \\
0.7 & $60^\circ$ & 0.0045 & 0.0039 & 0.0044
\end{tabular}
\end{ruledtabular}
\end{table*}

\begin{table*}[t]
\caption{\label{tab:flux_comparison2} Same as Table~\ref{tab:flux_comparison}, but for $\nu=10^{-4}$.}
\centering
\begin{ruledtabular}
\begin{tabular}{ccS[table-format=1.3]S[table-format=1.3]S[table-format=1.3]}
$e$ & $\theta_{\text{inc}}$ & {$\Delta\langle dE/dt \rangle\%$} & {$\Delta\langle dL_z/dt \rangle\%$} & {$\Delta\langle dQ/dt \rangle\%$} \\
\hline
0.1 & $20^\circ$ & 0.0040 & 0.0031 & 0.0058 \\
0.1 & $60^\circ$ & 0.0043 & 0.0030 & 0.0056 \\
0.3 & $20^\circ$ & 0.0104 & 0.0078 & 0.0110 \\
0.3 & $60^\circ$ & 0.0117 & 0.0087 & 0.0116 \\
0.5 & $20^\circ$ & 0.0216 & 0.0172 & 0.0213 \\
0.5 & $60^\circ$ & 0.0251 & 0.0201 & 0.0239 \\
0.7 & $20^\circ$ & 0.0372 & 0.0312 & 0.0371 \\
0.7 & $60^\circ$ & 0.0452 & 0.0385 & 0.0437
\end{tabular}
\end{ruledtabular}
\end{table*}

\begin{table*}[t]
\caption{\label{tab:flux_comparison3} Same as Table~\ref{tab:flux_comparison}, but for $\nu=10^{-3}$.}
\centering
\begin{ruledtabular}
\begin{tabular}{ccS[table-format=1.3]S[table-format=1.3]S[table-format=1.3]}
$e$ & $\theta_{\text{inc}}$ & {$\Delta\langle dE/dt \rangle\%$} & {$\Delta\langle dL_z/dt \rangle\%$} & {$\Delta\langle dQ/dt \rangle\%$} \\ 
\hline
0.1 & $20^\circ$ & 0.0404 & 0.0306 & 0.0583 \\
0.1 & $60^\circ$ & 0.0425 & 0.0303 & 0.0559 \\
0.3 & $20^\circ$ & 0.1043 & 0.0782 & 0.1100 \\
0.3 & $60^\circ$ & 0.1166 & 0.0868 & 0.1156 \\ 
0.5 & $20^\circ$ & 0.2160 & 0.1715 & 0.2132 \\
0.5 & $60^\circ$ & 0.2507 & 0.2004 & 0.2386 \\
0.7 & $20^\circ$ & 0.3708 & 0.3116 & 0.3698\\
0.7 & $60^\circ$ & 0.4502 & 0.3833 & 0.4350 \\
\end{tabular}
\end{ruledtabular}
\end{table*}



The convergence of the mode sums is exponential in $l$ and algebraic in $|k|$ and $|n|$. For a typical orbit with a requested flux accuracy of $\varepsilon=10^{-6}$, we require $l_{\max}\sim10$--$15$. The number of required $(k,n)$ pairs increases with eccentricity and inclination; an orbit with $e=0.7$ may require radial indices up to $|n|\simeq50$ and polar indices up to $|k|\simeq10$. The use of the analytic JH solver for the homogeneous solutions reduces the computation time per $(l,m,k,n)$ mode by approximately an order of magnitude compared with direct integration of the Sasaki--Nakamura equation, making the calculation of the necessary thousands of modes computationally feasible.

The EOB--Teukolsky model bridges the gap between analytic, mass-ratio-informed orbital dynamics and high-accuracy black hole perturbation theory. Its development has direct implications for space-based GW astronomy.
The present results should be interpreted within the adiabatic accuracy of the model. The orbital constants evolve through orbit-averaged dissipative fluxes, while conservative first-order self-force corrections and post-adiabatic effects are not yet included. These corrections shift the orbital frequencies at $\mathcal{O}(\nu)$ and can generate additional secular phase differences over a long inspiral~\cite{Pound:2005prd}. The calculations above therefore establish the behavior and computational viability of the EOB--Teukolsky framework at its current order, while also identifying the extensions required for precision parameter-estimation applications. Further developments will include conservative frequency shifts, the spin of the secondary object, and the plunge and ringdown phases. 

\section{Conclusion}
\label{sec:conclusion}

We present a full relativistic waveform model for EMRIs by synergistically combining the EOB formalism with the Teukolsky black hole perturbation theory. The EOB framework provides analytic, mass-ratio-informed solutions for generic bound geodesics within a deformed Kerr spacetime, while the Teukolsky formalism furnishes exact expressions for the gravitational radiation and its backreaction. This integration establishes a robust bridge between analytical dynamics and perturbative wave generation.


Key achievements of this work include: Successfully integrating EOB orbital dynamics with the solution of the Teukolsky equation to establish a complete waveform generation framework; by incorporating explicit mass-ratio dependence, we have computed the resulting differences in orbital energy fluxes. Furthermore, we have performed rigorous matching analyses on the final waveforms to ensure physical consistency and accuracy across different regimes; we have developed a highly efficient and stable numerical realization of this framework. The framework has been rigorously validated against established results across all relevant limiting cases, ensuring its reliability for production-level analysis.

This EOB-Teukolsky model delivers a significant framework in the accuracy of EMRI waveform templates in the strong-field regime with the mass-ratio correction, directly addressing the needs of future space-based GW observatories like LISA, Taiji and Tian-Qin. While the present work focuses on the EMRI regime, the EOB-based formulation may provide a useful starting point for future investigations of intermediate-mass-ratio inspirals, where finite-mass-ratio effects become increasingly important. We would like to point out that the accuracy of EOB in the extreme mass-ratio limit does not get complete validation with self-force. However, the EOB dynamics in principe can be callibrated to get more precise results based on future self-force data, therefore our EOB-Teukolsky framework can be improved automaticly.

Finally, because the waveform generation involves tens to hundreds of thousands of harmonics, the runtime depends sensitively on the available parallel computing resources. Future work will explore GPU acceleration and further optimization, with the long-term goal of producing full relativistic year-long EMRI waveforms on timescales suitable for large-scale data analysis.

\section*{Acknowledgments}
This work was supported by the National Key R\&D Program of China (Grant No. 2021YFC2203002), the National Science and Technology Major Project (No. 2024ZD1100601), the National Natural Science Foundation of China (Grant No. 12173071) and the China Postdoctoral Science Foundation (Grant NO. 2025M773434). This work made use of the High Performance Computing Resource in the Core Facility for Advanced Research Computing at Shanghai Astronomical Observatory.

\bibliographystyle{apsrev4-2}  
\bibliography{ref} 

\end{document}